\documentclass[aps,prl]{revtex4}
\usepackage[dvips]{graphicx}

\begin{document}

\title{Lattice Gauge Fixing and the Violation of Spectral Positivity}

\author {Christopher A. Aubin}
\author{Michael C. Ogilvie}
\affiliation{
Dept. of Physics
Washington University
St. Louis, MO 63130}

\begin{abstract}

Spectral positivity is known to be violated by some forms of
lattice gauge fixing. The most notable example is lattice Landau gauge,
where the effective gluon mass is observed to rise rather than fall with
increasing distance. We trace this violation to the use of quenched
auxiliary fields in the lattice gauge fixing process, and show that
violation of spectral positivity is a general feature of quenching. We
illustrate this with a simple quenched mass-mixing model in continuum field
theory, and with a quenched form of the Ising model. For lattice gauge
fixing associated with Abelian projection and lattice Landau gauge, we show
that spectral positivity is violated by processes similar to those found in
quenched QCD. For covariant gauges parametrized by a gauge-fixing parameter $%
\alpha $, the $SU(2)$ gluon propagator is well described by a simple
quenched mass-mixing formula. The gluon mass parameter appears to be
independent of $\alpha $ for sufficiently large $\alpha $.

\end{abstract}

\maketitle

Although many observables can be determined in lattice gauge theories
without gauge fixing, there are several reasons why gauge fixing
is desirable in lattice simulations. 
Gauge fixing is necessary to make the connection between continuum
and lattice gauge fields.
Continuum theories
of the origin of confinement often make predictions about the gauge field
propagator. Gauge fixing has also been a key technique in lattice studies
of confinement as well.\cite{Greensite:2003bk}
Important properties of the quark-gluon plasma phase of QCD,
such as screening masses,
are contained in the finite-temperature gluon propagator.

Techniques for lattice gauge fixing have been known for
some time.\cite{Mandula} It has been clear from the beginning
that non-Abelian lattice gauge field propagators show
a violation of spectral positivity.
This is readily seen from the
effective mass: for a normal operator which connects only states of
positive norm to the vacuum, the effective mass monotonically decreases with
distance to the lightest mass state coupling to the operator. Covariant
gauge gluon propagators have an effective mass increasing with distance. In
one sense, this is not surprising. We know from perturbation theory that
covariant gauges contain states of negative norm. However, that knowledge
has neither explained the form of the lattice gluon propagator nor aided in the
interpretation of the mass parameters measured from it. In fact, no
similar violation of spectral positivity is observed in the 
$U(1)$ case \cite{Coddington:yz},
which has negative-norm states in covariant gauges.

In lattice simulations, gauge fixing has typically involved choosing a
particular configuration on each gauge orbit. 
A brief review of this approach
is given in Ref.~\cite{Mandula:nj}.
In the continuum, on the other hand, 
gauge fixing usually includes a parameter that 
causes the functional integral to peak around a
particular configuration on the gauge orbit. 
As shown below, the extension of this idea to lattice gauge theories
makes clear that lattice gauge fixing is a form of quenching,
with the gauge transformations acting as quenched fields. 
As has been demonstrated in
quenched QCD, quenching can violate spectral positivity, with significant
effects on many observables.\cite{Golterman:1994jr,Bardeen:2001yz}

We begin with a review of lattice gauge fixing, including the generalization
of lattice Landau gauge to covariant gauges 
with a gauge parameter.\cite{Zwanziger:tn,Parrinello:1990pm,Fachin:1991pu} 
This generalization will be directly interpreted as a quenched Higgs
theory. We then explore the origin of violations of spectral positivity
in some simple lattice and continuum models of quenching. 
Simulation results for the effective mass of an $SU(2)$ lattice gauge field
will show behavior very similar to these models as the gauge fixing 
parameter is varied.
We will argue that spectral positivity violations in both lattice
covariant gauges and in studies of Abelian projection originate
in the quenching process. 

The standard approach to lattice gauge fixing is a two 
step process.\cite{Mandula:nj}
An ensemble of
lattice gauge field configurations is generated using standard Monte Carlo
methods, corresponding to a
functional integral 
\begin{equation}\label{eq:gen_functional}
  Z_{U}=\int \left[ dU\right] e^{S_{U}\left[ U\right] } \ ,
\end{equation}
where $S_{U}$ is a
gauge-invariant action for the gauge fields, \textit{e.g.}, the Wilson
action.
The gauge action $S_{U}$ is
invariant under gauge transformations of the form 
$U_{\mu }\left( x\right) \rightarrow g\left( x\right) U_{\mu }\left( x\right)
  g^{+}\left( x+\mu \right)$. 

In order to measure gauge-variant observables, each field configuration in
the $U$-ensemble may be placed in a particular gauge, \textit{i.e.}, a gauge
transformation is applied to each configuration in the $U$-ensemble which
moves the configuration along the gauge orbit to a gauge-equivalent
configuration satisfying a lattice gauge fixing condition. The simplest
gauge choice is defined by maximizing 
$ \sum_{x,\mu }Tr\,\left[ U_{\mu }\left( x\right) +U_{\mu }^{+}\left( x\right)
    \right]$
for each configuration over the class of all gauge transformations.
Any local extremum of this functional
satisfies a lattice form of the Landau gauge condition: 
\begin{equation}\label{eq:Landau_condition}
  \sum_{\mu }\left[ A_{\mu }\left( x+\mu \right) -A_{\mu }\left( x\right)
  \right] =0 
\end{equation}
where $A_{\mu }\left( x\right) $ is a lattice approximation to the continuum
gauge field, given by 
\begin{equation}\label{eq:lattice_Amu}
  A_{\mu }\left( x\right) =\frac{U_{\mu }\left( x\right) -U_{\mu }^{+}\left(
    x\right) }{2i}-\frac{1}{N}Tr\left[ \frac{U_{\mu }\left( x\right) -U_{\mu
      }^{+}\left( x\right) }{2i}\right] \text{.} 
\end{equation}
Other gauge-fixing conditions may also be used
\cite{Giusti:2001xf}, and lattice improvement techniques 
can be applied to the definition of
$A_{\mu }$ to reduce discretization errors as well. 
The global maximization needed is often
implemented as a local iterative maximization. The issue of Gribov
copies arises in lattice gauge fixing because such a local algorithm
tends to find local maxima of the gauge-fixing functional. There are
variations on the basic algorithm that ensure a unique choice from
among local maxima.\cite{Giusti:2001xf}

For analytical purposes, it is necessary to generalize this procedure
\cite{Fachin:1991pu}, so that a given single configuration of gauge
fields will be associated with an ensemble of configurations of
$g$-fields. We will generate this ensemble using
\begin{equation}\label{eq:gaugefix_action}
  S_{gf}\left[ U,g\right] =\sum_{l}\frac{\alpha }{2N}Tr\,\left[ g\left(
    x\right) U_{\mu }\left( x\right) g^{+}\left( x+\mu \right) +g\left( x+\mu
    \right) U_{\mu }^{+}\left( x\right) g^{+}\left( x\right) \right] 
\end{equation}
as a weight function to select an ensemble of $g$-fields.
The sum over $l$ is a sum over all links of the lattice.
The normal
gauge-fixing procedure is formally regained in the limit $\alpha \rightarrow
\infty $. Computationally, this can be implemented as a Monte Carlo
simulation inside a Monte Carlo simulation. 

Note that the $g$-fields must
be thought of as quenched variables, since they do not affect the
$U$-ensemble.
The expectation value of an observable $O$, gauge-invariant or not, is
given by 
\begin{equation}\label{eq:new_exp_O}
  \left\langle O\right\rangle =\frac{1}{Z_{U}}\int \left[ dU\right]
  \,e^{S_{U}\left[ U\right] }\frac{1}{Z_{gf}\lbrack U\rbrack }\int \left[
    dg\right] e^{S_{gf}\left[ U,g\right] }\,O  \ ,
\end{equation}
where 
\begin{equation}\label{eq:gen_fun_gf}
  Z_{gf}\lbrack U\rbrack =\int \left[ dg\right] e^{S_{gf}\left[ U,g\right] }%
  \text{.} 
\end{equation}
Formally, the field $g$ is a quenched scalar field with two
independent symmetry groups, $G_{global}\otimes G_{local}$, so that it
appears to be in the adjoint representation of the gauge group, but the left
and right symmetries are distinct. The generating functional $Z_{gf}\lbrack
U\rbrack $ is in some ways a lattice analog of the inverse of the Fadeev-Popov
determinant.\cite{Bock:2000cd} However, there are important
differences. Note immediately that $Z_{gf}\lbrack U\rbrack $ depends on the
gauge-fixing parameter $\alpha $. More fundamentally, 
the lattice formalism resolves the Gribov
ambiguity. By construction, gauge-invariant observables are evaluated by
integrating over all configurations. Gauge-variant quantities receive
contributions from Gribov copies, always with positive weight. 
Thus the connection between
this formalism for
lattice gauge fixing and gauge fixing in the continuum is not simple.
Furthermore, alternative lattice gauge fixing procedures have been proposed,
along with new gauge choices specific to the lattice. A comprehensive
review of gauge fixing technology is available.\cite{Giusti:2001xf}

We begin our analysis of spectral positivity violation
with the simplest
model of quenching possible: 
two free, real scalar fields with a non-diagonal mass
matrix. The Lagrangrian is

\begin{equation}\label{eq:mixing_L}
  L=\frac{1}{2}\left[ \left( \partial \phi _{1}\right) ^{2}+m_{1}^{2}\phi
    _{1}^{2}\right] +\frac{1}{2}\left[ \left( \partial \phi _{2}\right)
    ^{2}+m_{2}^{2}\phi _{2}^{2}\right] -\mu ^{2}\phi _{1}\phi _{2} \ .
\end{equation}
We treat the quenched approximation of this model in a manner completely
parallel to our discussion of lattice gauge fixing above. We divide the
action into three parts 
$  S=S_{1}+S_{2}+S_{12}$,
where $S_{1}$ and $S_{2}$ are functionals only of $\phi_{1}$ and
$\phi_{2}$, respectively, and $S_{12}$ contains the mixing term. We quench the
field $\phi _{2}$. Although there are no loops in this simple theory,
quenching implies that $\phi _{2}$ cannot appear as an internal line in
the complete propagators. 
The generating functional
in the
quenched approximation, including sources $J_{1}$ and $J_{2}$ is 
\begin{equation}\label{eq:mixing_Z}
  Z=\int \left[ d\phi _{1}\right] e^{-S_{1}+\int J_{1}\phi _{1}}\frac{\int
    \left[ d\phi _{2}\right] e^{-S_{2}-S_{12}+
      \int J_{2}\phi _{2}}}{\int \left[ d
      \widetilde{\phi }_{2}\right] e^{-S_{2}-S_{12}}}\text{.} 
\end{equation}
where we have introduced a kind of ghost variable $\widetilde{\phi }_{2}$;
space-time variables are implicit.

\begin{figure}
  \includegraphics[width=3in]{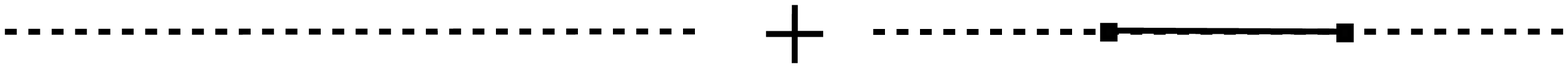}
  \caption{Exact propagator for the quenched $\phi_2$ field in the simple mass
    mixing model.}
\label{fig:phi2phi2}
\end{figure}

From the generating functional
 we can obtain the $\left\langle \phi _{1}\phi
_{1}\right\rangle $ and $\left\langle \phi _{2}\phi _{2}\right\rangle $
propagators. In momentum space, the $\left\langle \phi _{1}\phi
_{1}\right\rangle \,$propagator is $1/(p^{2}+m_{1}^{2})$, since $\phi _{1}$
is unaffected by $\phi _{2}$. On the other hand, the $\left\langle \phi
_{2}\phi _{2}\right\rangle $ propagator is 
\begin{equation}\label{eq:phi2_prop}
  \frac{1}{p^{2}+m_{2}^{2}}+\frac{1}{p^{2}+m_{2}^{2}}\mu ^{2}\frac{1}{%
    p^{2}+m_{1}^{2}}\mu ^{2}\frac{1}{p^{2}+m_{2}^{2}}\text{.} 
\end{equation}
An alternative diagrammatic procedure is to sum Dyson's series, as
shown in Fig.~\ref{fig:phi2phi2}, noting that
the $\left\langle \phi _{2}\phi _{2}\right\rangle $ propagator is truncated
at two terms. 
The propagator has a structure similar to
the $\eta'$ propagator in quenched QCD
\cite{Golterman:1994jr,Bardeen:2001yz}; the $\eta'$ has 
a double pole form in quenched QCD
when singlet self-energy graphs are approximated by a constant.
The $\left\langle \phi _{2}\phi _{2}\right\rangle $ propagator
also may be written as 
\begin{equation}\label{eq:phi2_prop_2}
  \left( 1-\frac{\mu ^{4}}{\left( m_{2}^{2}-m_{1}^{2}\right) ^{2}}\right)
  \allowbreak \frac{1}{p^{2}+m_{2}^{2}}+\frac{\mu ^{4}}{\left(
    m_{2}^{2}-m_{1}^{2}\right) ^{2}}\frac{1}{p^{2}+m_{1}^{2}}+\frac{\mu ^{4}}{%
    m_{1}^{2}-m_{2}^{2}}\allowbreak \frac{1}{\left( p^{2}+m_{2}^{2}\right) ^{2}}%
  \text{.} 
\end{equation}
This propagator always violates spectral positivity because of the double
pole term, $1/\left( p^{2}+m_{2}^{2}\right) ^{2}$, which has a coefficient
whose sign depends on $m_{1}^{2}-m_{2}^{2}$.
Another possible violation of spectral positivity occurs for sufficiently
strong mixing: if $\mu ^{4}>\left( m_{2}^{2}-m_{1}^{2}\right) ^{2}$, there
is a simple pole at $p^{2}=-m_{2}^{2}$ with negative residue.

The form of the $\left\langle \phi _{2}\phi _{2}\right\rangle $
propagator in coordinate space is very interesting, and forms the
basis for our study of other quenched theories.  In any number of
dimensions, we can consider propagators using wall sources,
\textit{i.e.}, of co-dimension $1$. This has the effect of setting the
momentum equal to zero in all the directions of the wall. For wall
sources, we have the propagator
\begin{eqnarray}\label{eq:coor_space_prop}
  G(x) &=&\left( 1-\frac{\mu ^{4}}{\left( m_{2}^{2}-m_{1}^{2}\right) ^{2}}%
  \right) \allowbreak \frac{1}{2m_{2}}e^{-m_{2}\left| x\right| }+\frac{\mu ^{4}%
  }{\left( m_{2}^{2}-m_{1}^{2}\right) ^{2}}\frac{1}{2m_{1}}e^{-m_{1}\left|
    x\right| } \nonumber\\
  &&+\frac{\mu ^{4}}{m_{1}^{2}-m_{2}^{2}}\frac{1}{4m_{2}^{3}}e^{-m_{2}\left|
    x\right| }\left( 1+m_{2}\left| x\right| \right) \allowbreak \text{.}
\end{eqnarray}
The factor $m_{2}\left| x\right| e^{-m_{2}\left|x\right|}$
shows an initial rise rather than a decay with increasing $\left|x\right|$,
violating spectral positivity.


We define an effective mass associated with the $\phi _{2}$ field as 
\begin{equation}\label{eq:eff_mass}
  m_{eff}(x)=\lim_{a\rightarrow 0}\frac{1}{a}\ln \left( \frac{G(x)}{G(x+a)}%
  \right) =-\frac{d}{dx}\ln (G(x))\text{.} 
\end{equation}
One can easily check explicitly 
that $m_{eff}(x)\rightarrow \min (m_{1},m_{2})\,$\ as $%
x\rightarrow \infty $.
For any field theory which obeys spectral positivity, $%
m_{eff}(x)$ monotonically decreases to its limiting value.
Theories violating spectral positivity may display a complicated
behavior for $m_{eff}(x)$ before the eventual onset of asymptotic behavior.

\begin{figure}
  \includegraphics[width=6in]{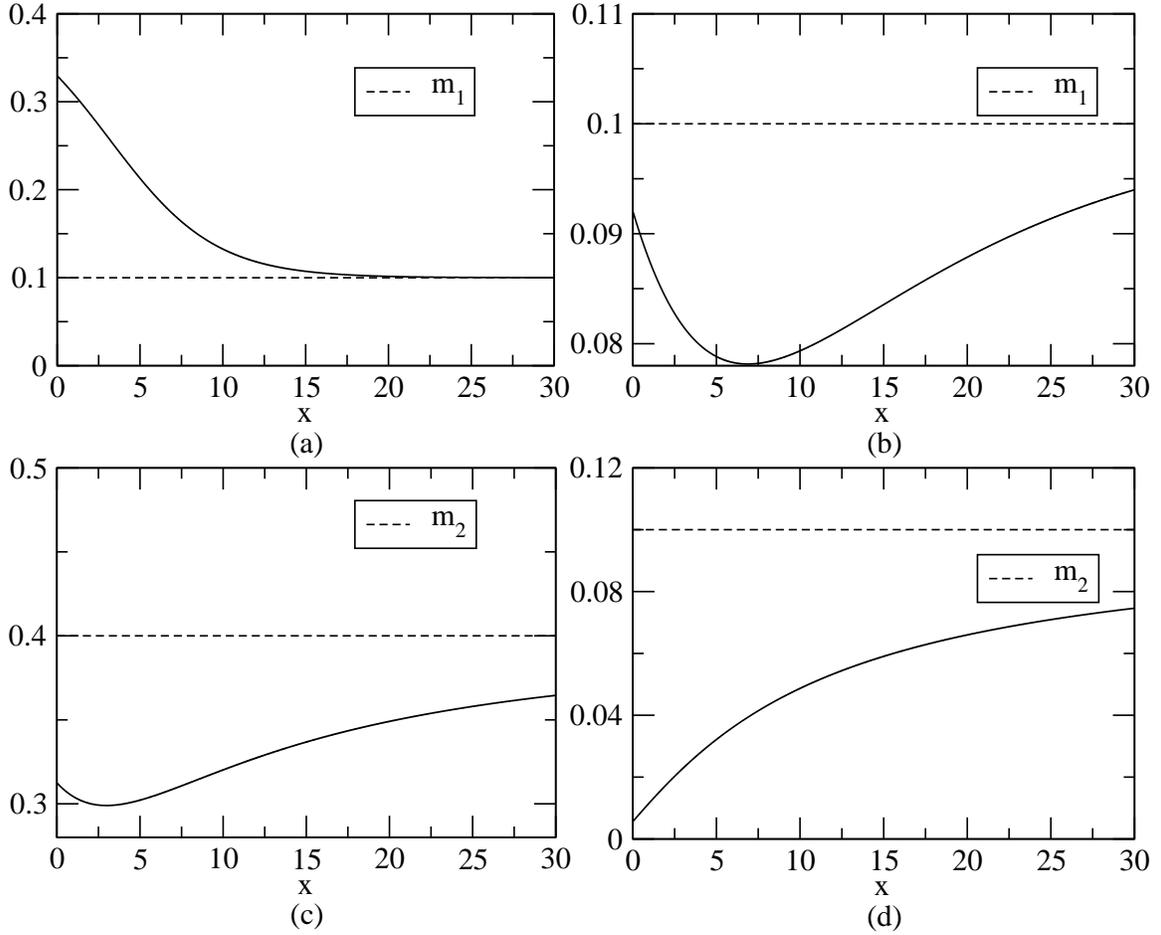}
  \caption{The effective mass associated with $\phi_2$ as a function
  of $x$. The parameters used are: 
  (a) $m_1=0.1$, $m_2=0.4$, $\mu=0.2$;
  (b) $m_1=0.1$, $m_2=0.2$, $\mu=0.2$;
  (c) $m_1=0.5$, $m_2=0.4$, $\mu=0.4$;
  (d) $m_1=0.5$, $m_2=0.1$, $\mu=0.4$.}
  \label{fig:YY}
\end{figure}

We have identified three different possible behaviors for $m_{eff}(x)$
in this simple quenched model. If the mixing parameter $\mu $ is
sufficiently small and $m_{1}<m_{2}$, $m_{eff}(x)$ monotonically
decreases to its value at infinity, as in a normal field theory which
obeys spectral positivity, as shown in Fig.~\ref{fig:YY}(a). 
As $\mu$ is increased relative to $m_1$ and $m_2$,
$m_{eff}(x)$ may develop a minimum,
as displayed in Fig.~\ref{fig:YY}(b).
On the other hand, if $m_{2}<m_{1}$,
the behavior seen in Fig.~\ref{fig:YY}(a)
is not possible, and only the behaviors seen
in Fig.~\ref{fig:YY}(c) and Fig.~\ref{fig:YY}(d) are possible.
In Fig.~\ref{fig:YY}(d), 
the minimum has moved to $x=0$.
For sufficiently small $\mu $, these effects
are difficult to observe, and $m_{eff}(x)$ is essentially
equal to $m_{2}$ for all $x$. 
Regardless of the relative size of $m_1$ and $m_2$,
an observable
violation of spectral positivity associated with $m_{eff}(x)$ not
monotonically decreasing indicates a significant mixing parameter $\mu$.

Similar behavior can be observed in 
a very simple lattice model based on the Ising model, where
real-space arguments can be used to find an approximate propagator.
We consider two coupled one-dimensional Ising models, with spins $\mu _{i}$,
 $\sigma _{i}\in \left\{ -1,+1\right\} $ and respective nearest-neighbor
couplings $J$ and $K$. The $\sigma$ spins are coupled to the $\mu $ spins
via an interaction of the form $\sum_{i}L\sigma _{i}\mu _{i}$, and the $%
\sigma $'s are quenched. 
This simple model is a form of spin glass, with the averaging over the ensemble
of $\mu$ spins representing the ``quenching'' process. 

The $\sigma \,$propagator is given by 
\begin{equation}\label{eq:sigma_prop_initially}
  \left\langle \sigma _{0}\sigma _{n}\right\rangle =\frac{1}{Z_{\mu }}%
  \sum_{\left\{ \mu \right\} }\exp \left[ \sum_{i}J\mu _{i}\mu _{i+1}\right] 
  \frac{1}{Z_{\sigma }\left[ \mu \right] }\sum_{\left\{ \sigma \right\}
  }\sigma _{0}\sigma _{n}\exp \left[ \sum_{i}\left( K\sigma _{i}\sigma
    _{i+1}+L\sigma _{i}\mu _{i}\right) \right] \ ,
\end{equation}
where $Z_\mu$ is the partition function for $\mu$ and $Z_\sigma[\mu]$ is
the partition function for $\sigma$ in the presence of a particular $\mu$
background.
The parameter $L\,$is a mixing parameter. We can approximately evaluate the $%
\sigma \,$propagator for $J$, $K$, and $L$ sufficiently small
by considering the direct contribution $\left( \tanh
K\right) ^{n}$ combined with mixing of $\sigma \,$with $\mu $. This indirect
term can be written as (compare Fig.~\ref{fig:phi2phi2})
\begin{equation}\label{eq:indirect_term}
  \sum_{p=1}^{n}\sum_{m=0}^{n-p}\left( \tanh K\right) ^{n-p}
  \left( \tanh L \right)^{2} \left(
  \tanh J\right) ^{p}.
\end{equation}
After performing the summations, the propagator is given
approximately as
\begin{equation}\label{eq:sigma_prop}
  \left\langle \sigma _{0}\sigma _{n}\right\rangle \approx 
  \left( \tanh K\right)
  ^{n}+\left( \tanh L\right) ^{2}\left( \tanh K\right) ^{n}\left[ \frac{nx}{%
      \left( 1-x\right) }-\frac{x^{2}(1-x^{n})}{\left( 1-x\right)
  ^{2}}\right] \ .
\end{equation}
where $x=\allowbreak $ $\tanh J/\tanh K$.
The $n(\tanh K)^{n}$ factor signals a violation of spectral
positivity, just as the
$m \left| x\right| \exp \left( -m\left| x\right|\right) $
 term did in the mixing model. Of course, the arguments which
led to Eq.~(\ref{eq:phi2_prop}) and Eq.~(\ref{eq:sigma_prop}) are
essentially the same, but carried out in momentum space and real
space, respectively. For small $J$, $ K $, and $L$, 
Eq.~(\ref{eq:sigma_prop}) fits lattice simulations of the 
$\left\langle \sigma _{0}\sigma _{n}\right\rangle $ propagator well.

\begin{figure}
  \includegraphics[width=6in]{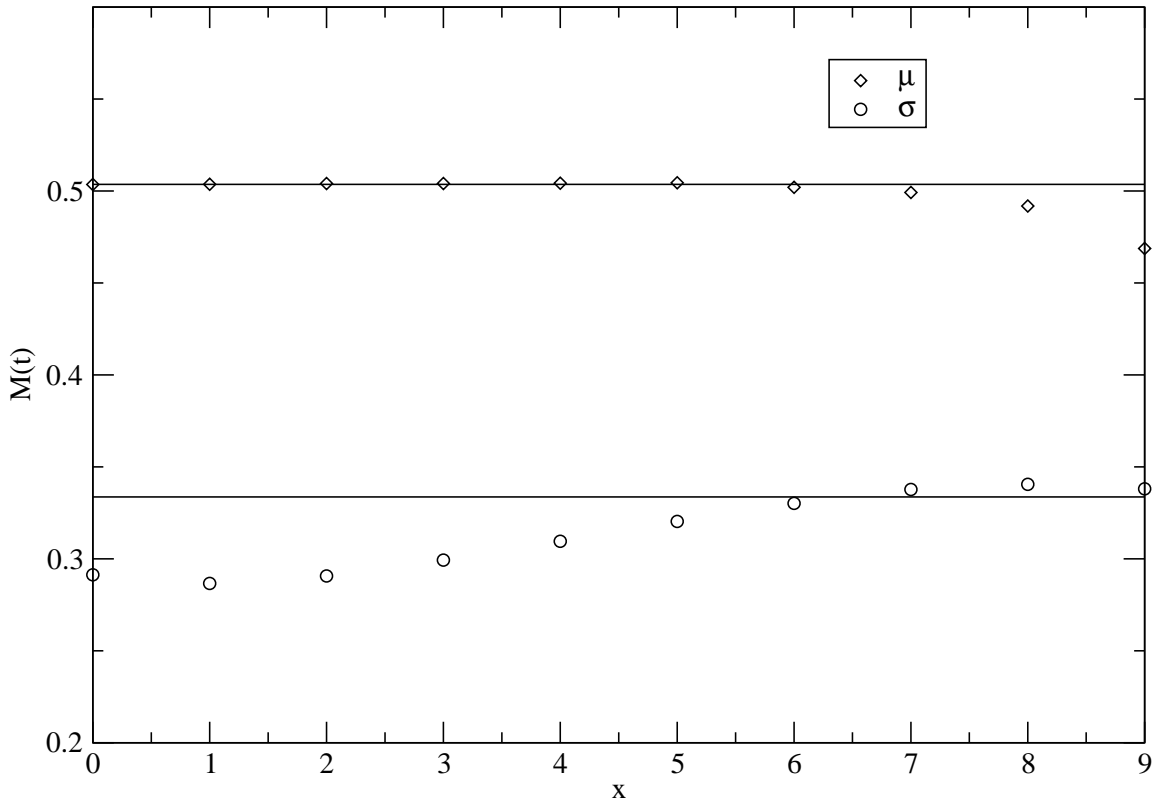}
  \caption{Effective masses for $\sigma$ and $\mu$ in the $1-d$ Ising
  model.}\label{fig:ising}
\end{figure}

In Fig.~\ref{fig:ising}, we show the effective mass determined from
the $\left\langle \sigma _{0}\sigma _{n}\right\rangle $ and
$\left\langle \mu _{0}\mu _{n}\right\rangle $ propagators for the
parameter set $J=0.7$, $K=0.9$, and $L=0.3$ for a one-dimensional
lattice of size $26$.  The propagators were obtained from $40000$ heat
bath sweeps of the $\mu$ variables; after each such sweep, $100$ heat
bath sweeps of the $\sigma$ variables were carried out.
The parameters
$K$ and $L$ were chosen empirically so as to display a clear violation
of spectral positivity.
The $\mu $
mass fits very well with the analytical solution $m=-\ln \tanh (J)$
for the $d=1$ Ising model out to a distance of $8$.
Note that the $\sigma $ reaches its
asymptotic value of $-\ln (\tanh K)$ from below, and only
at $n\simeq 8$.
The similarity to the simple field theoretic model of quenching is
clear.

We will now show that the $SU(2)$ lattice gluon propagator regarded as
a function of $\alpha$ shows behavior similar to that of the other,
simpler quenched models studied above.  Simulations of this type of
lattice field theory, with stochastic quenched gauge fixing fields,
were first performed by Henty \textit{et al.} \cite{Henty:1996kv}, who
studied the case of $SU(3)$ as a function of $\alpha $ at $\beta=5.7$
on $8^4$ lattices.  They found evidence for a first-order phase
transition as $\alpha$ was varied, but did not determine the full
phase diagram in the $\alpha$-$\beta$ plane.  They also found that the
gluon propagator was dependent on $\alpha$, a result which could
be anticipated from
the strong-coupling expansion.\cite{Fachin:1991pu}

Let us consider for the moment the unquenched version of the gauge
fixing model. This is a model with scalar fields
in the fundamental representation of the gauge group
in addition to the gauge fields.
The scalar
fields explicitly break the $Z(N)$ global symmetry associated with
confinement in the pure gauge case, and external color charges are
screened.  As first shown by Fradkin and Shenker
\cite{Fradkin:1978dv}, this leads to a connection between
the strong-coupling, confining phase and the Higgs phase, so the
two phases are not actually distinct.  We have verified that this
phase structure is preserved in the quenched form of the model.  For
$\beta$ sufficiently large, there is a line of first-order phase
transitions in the $\beta$-$\alpha$ plane. It is very reasonable that
such a line exists in the quenched model, since it can be thought of as
the continuation of the critical point of a pure spin model at
$\beta=\infty$.  However, this line terminates at a critical end
point; for sufficiently small $\alpha$, the nominal confining phase
($\beta$ small) and Higgs phase ($\beta$ large) are directly
connected.  This observation forms the starting point for a detailed
analysis of the model.\cite{us_longer_ms}

\begin{figure}
  \includegraphics[width=6in]{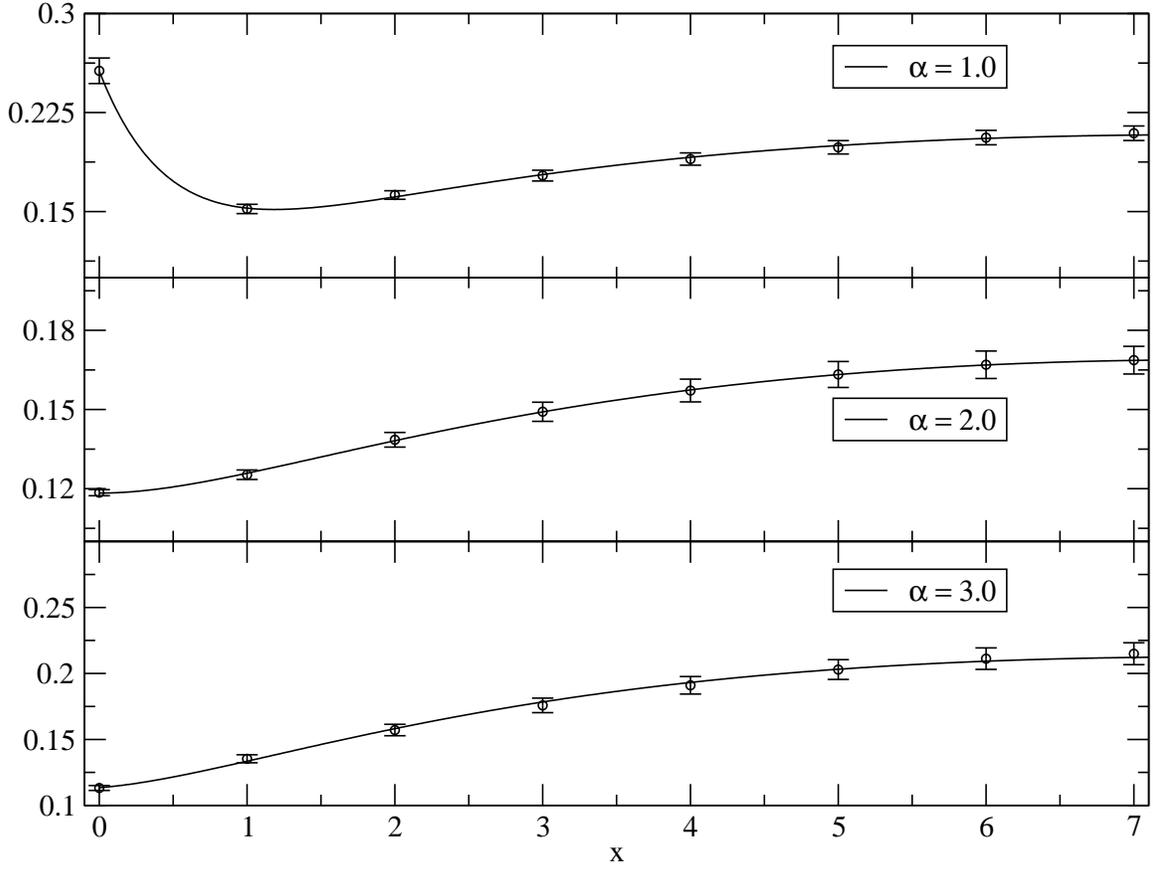}
  \caption{Effective masses for three values of the gauge fixing 
    parameter, $\alpha = 1.0$, $2.0$ and $3.0$.}
   \label{fig:three_masses}
\end{figure}

We have performed simulations of $SU(2)\,$gauge theory at $\beta =2.6$
with $\alpha $ ranging from $1.0$ to $3.0$ on a $12^{3}\times 16$
lattice.  At this value of $\beta$, there is a first-order phase
transition at $\alpha \approx 0.83$.  We have fit the data using
a simple generalization of the quenched mixing model.  The coordinate
space propagator has the form
\begin{equation}\label{eq:mixing_generalization_prop}
  \left( A+Bm_{2}\left| x\right| \right)e^{-m_{2}\left|x\right| }
   + C e^{-m_{1}\left| x\right| }\text{.} 
\end{equation}
This form for the propagator follows from the replacement of the
mixing parameter $\mu^4$ in Eq.~(\ref{eq:phi2_prop})
by the more general form $\mu^2 (p^2+m_3^2)$.

\begin{figure}
  \includegraphics[width = 6in]{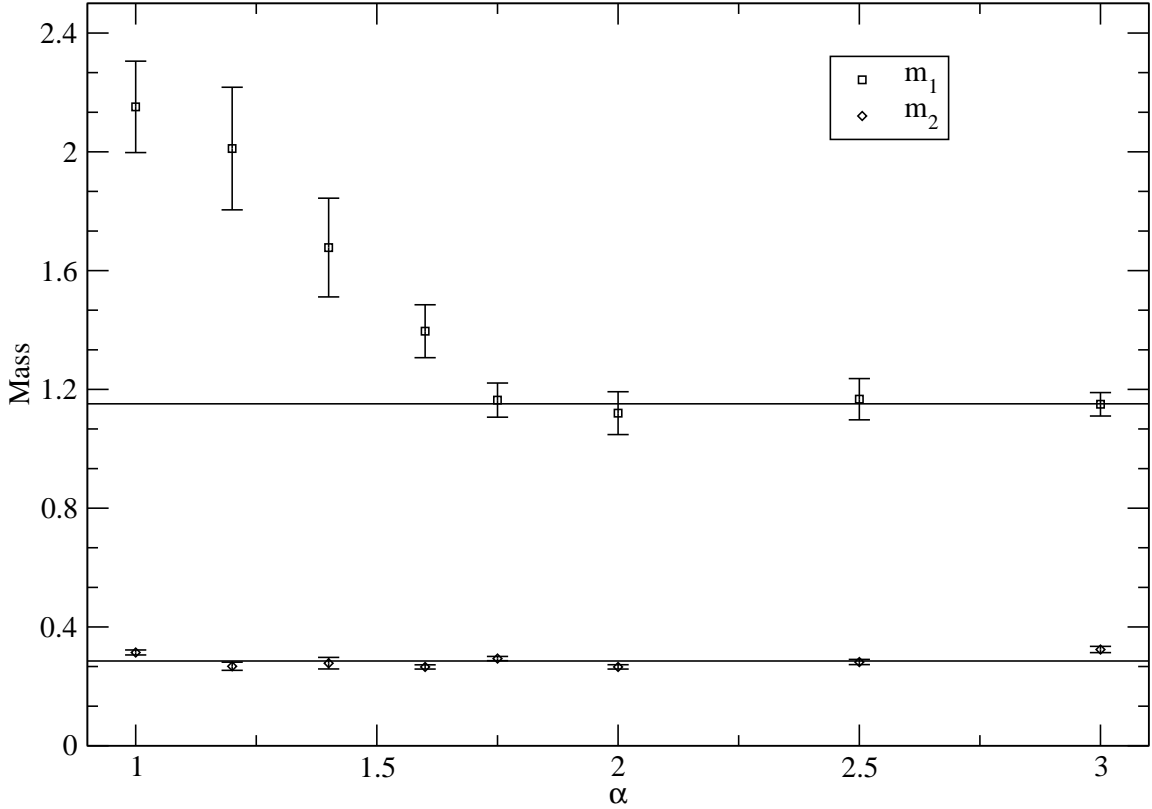}
  \caption{Values of the two mass parameters in 
    eq.~(\ref{eq:mixing_generalization_prop}) as a
    function of $\alpha$. The light mass, $m_2$ 
    is approximately constant at the
    value of $0.283(7)$, while the heavy mass initially decreases with
    increasing $\alpha$, reaching a constant value of $1.151(8)$.}
\label{fig:m1m2}
\end{figure}

In Fig.~\ref{fig:three_masses}, we show the effective mass as a
function of $x\,$for $\alpha =1$, $2$, and $3$.
The solid lines are obtained from fits to the propagator using
Eq.~(\ref{eq:mixing_generalization_prop}). 
The similarity to the other quenched models
is quite clear. 
For small
$\alpha$ there is an initial decrease and then rise of the
effective mass, much like Fig.~\ref{fig:YY}(c); as $\alpha$
increases, this minimum vanishes.
We plot in Fig.~\ref{fig:m1m2} the best-fit
values of $m_1$ and $m_2 $ as a function of $\alpha$.  Note that the
behavior of $m_2$ is consistent with it being constant in this region,
while $m_1$ appears to be decreasing to a constant limit as $\alpha$
increases.

Our results are roughly consistent with the work of Leinweber
\textit{et al.}, who performed high-precision studies of the $SU(3)$
gluon propagator at $\alpha =\infty $.\cite{Leinweber:1998uu} Among a
large variety of possible functional forms for the gluon propagator,
they found that their data was best fit by the functional form
\begin{equation}\label{eq:leinweber_form}
  G(k) = Z \left[ \frac{AM^{2\delta}}{(k^2 + M^2)^{1+\delta}}
      + \frac{1}{k^2 + M^2} L(k^2, M) \right] \ ,
\end{equation}
with $L(k^2, M)$ an infrared-regulated version of the asymptotic
behavior of the renormalized gluon propagator in the continuum.
Their best fit was achieved
with the parameters $\delta = 2.2^{+0.1+0.2}_{-0.2-0.3}$, $M =
(1020\pm 100\pm 25)\,\textrm{MeV}$, and $A = 9.8^{+0.1}_{-0.9}$.
Many other functional forms were ruled out.

Our results suggest that the lighter mass parameter $m_2$ is
independent of $\alpha $, at least for 
large $\alpha $ (in the Higgs phase).
If $m_2$ is indeed independent of gauge choice, as least
within the class of covariant gauges considered, it seems natural to
identify it as the gluon mass.  As a consequence of the quenched
character of lattice gauge fixing, this state partially mixes with
another, heavier state, with a mass on the order of the scalar or
vector glueball.\cite{Teper:1998kw}

Note that the value of the lightest mass in the propagator may be
difficult to extract from the effective mass. While it is true that
the effective mass tends asymptotically to the lightest mass, the
approach to the limit can be much slower than in a conventional field
theory obeying spectral positivity.  
For example, at $\alpha=3.0$, $m_{eff}$ at $x=7$
is substantially lighter than $m_2$.
Having a theoretical basis
for the form of the propagator is crucial in estimating the mass.

Another application of lattice gauge fixing is
Abelian projection, a method for investigating the confining
properties of gauge theories. 
In lattice gauge theories, Abelian projection is
implemented as an algorithm for extracting an ensemble of Abelian
gauge field configurations from an ensemble of non-Abelian
configurations. A notable success of lattice studies of Abelian
projection \cite{Suzuki:1989gp,Hioki:1991ai}
has been the correlation of
the string tension of the projected theory with the string tension of
the underlying non-Abelian theory.

These lattice studies of Abelian projection typically use gauge fixing
in an integral way. Taking for clarity the case of $SU(2)$, the gauge
fixing functional is
\begin{equation}\label{eq:abel_gf_functional}
  S_{gf}=\sum_{x,\mu }\frac{\alpha }{2}Tr\left[ g\left( x\right) U_{\mu
    }\left( x\right) g^{+}\left( x+\mu \right) \sigma _{3}g\left( x+\mu \right)
    U_{\mu }^{+}\left( x\right) g^{+}\left( x\right) \sigma
    _{3}\right] \ ,
\end{equation}
which is conventionally maximized over the gauge orbit, corresponding to
the limit $\alpha \to \infty$
in the formalism used here. 
An initial study of the phase structure in the $SU(2)$ case
finds evidence for a first order phase transition as
$\alpha$ is varied at $\beta=2.4$.\cite{Mitrjushkin:2001hr}
The aim of this
procedure is to transform an $SU(2)$ configuration into a gauge-equivalent one
which lies mostly in a given $U(1)$ subgroup. After this gauge-fixing, the
actual projection to $U(1)$ is performed. 

In the case where no gauge fixing is done ($\alpha=0$), 
and only projection occurs,
Faber \textit{et al.} \cite{Faber:1998en} and Ogilvie
\cite{Ogilvie:1998wu} have proved that the asymptotic string tension
measured in the projected and underlying theories are the same.
Furthermore, Ogilvie \cite{Ogilvie:1998wu} has proven that this result
should continue to hold for small $\alpha $, under the assumption that the
gauge fixing does not violate spectral positivity. However, the fact
that the string tension evaluated using various forms of
Abelian projection with gauge
fixing is consistently slightly different from
the actual non-Abelian string tension
\cite{Stack:2000kf,Bornyakov:2000cd}
suggests that a violation of spectral positivity may indeed be occurring.

\begin{figure}
  \includegraphics[width = 3in]{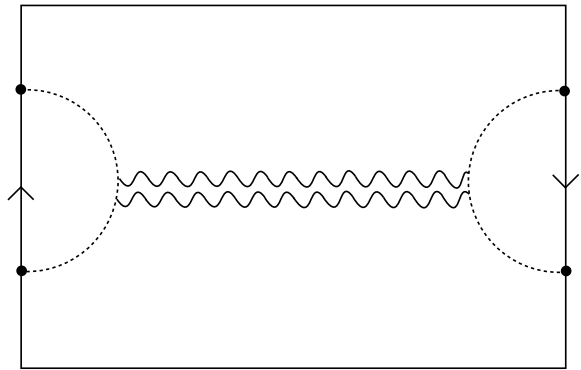}
  \caption{An example of a problematic diagram when calculating
  the expectaion value of Eq.~(\ref{eq:wilson_loop}). The solid line
  is the Wilson Loop, the dotted lines are $\phi$ propagators, and the
  wavy lines are gluons.}
\label{fig:AbPr}
\end{figure}

We identify the origin of this violation as the presence of a quenched
scalar field. 
The case of $SU(2)$ is particularly clear.
Note
that the combination $g^{+}\left( x\right) \sigma _{3}g\left( x\right)
$ occurring in $S_{gf}$ can be written as a Hermitian scalar field
$\phi (x)$, where $\phi $ transforms as the adjoint representation of
the gauge group.
The field $\phi$
is traceless, $Tr(\phi )=0$, and satisfies $Tr(\phi
^{2})=2$. The gauge fixing action is thus equivalent to an adjoint scalar
action of the form
\begin{equation}\label{eq:adj_scalar_action}
  S_{gf}=\sum_{x,\mu }\frac{\alpha }{2}Tr\left[ \phi \left( x\right) U_{\mu
    }\left( x\right) \phi \left( x+\mu \right) U_{\mu }^{+}\left( x\right)
    \right] \text{.} 
\end{equation}
As we have seen, such quenched fields naturally lead to violations of
spectral positivity.
Suppose we wish to measure a $U(1)$ projected Wilson loop. This may be
obtained from the expectation value of
\begin{equation}\label{eq:wilson_loop}
  Tr\, \prod_j \frac{1}{2}(1+\sigma_3) g_j U_j g^+_{j+1}
=Tr\, \prod_j \frac{1}{2}(1+\phi_j) U_j \ ,
\end{equation}
where the product is ordered along a closed path labeled by the index $j$.
The $U(1)$ projected loop is represented in the underlying quenched Higgs theory
as a sum of Wilson loops with all possible insertions of $\phi$ at lattice
sites on the path. When four or more $\phi$ fields are inserted, problematic subdiagrams
appear of the type shown in Fig. \ref{fig:AbPr}.
Such terms lead to a violation of spectral positivity:
there are no internal $\phi$ loops in the quenched
approximation, and an infinite set of diagrams occurring
in the full, unquenched theory is omitted.
This exactly parallels quenched QCD.

\acknowledgments{ This work was partially supported by the U.S. Department
of Energy under grant number DE-FG02-91ER40628.}


\begin{thebibliography}{30}

\bibitem{Greensite:2003bk}
J.~Greensite,
[hep-lat/0301023].

\bibitem{Mandula}
J. Mandula and  M. Ogilvie,
Phys. Lett. B185 (1987) 127.

\bibitem{Coddington:yz}
P.~Coddington, A.~Hey, J.~Mandula and M.~Ogilvie,
Phys.\ Lett.\ B {\bf 197}, 191 (1987).

\bibitem{Mandula:nj}
J.~E.~Mandula,
Phys.\ Rept.\  {\bf 315}, 273 (1999).

\bibitem{Golterman:1994jr}
M.~F.~Golterman,
Pramana {\bf 45}, S141 (1995)
[hep-lat/9405002].

\bibitem{Bardeen:2001yz}
W.~A.~Bardeen, A.~Duncan, E.~Eichten, N.~Isgur and H.~Thacker,
Nucl.\ Phys.\ Proc.\ Suppl.\  {\bf 106}, 254 (2002)
[hep-lat/0110187], Phys.\ Rev.\ D {\bf 65}, 014509 (2002)
[hep-lat/0106008].

\bibitem{Zwanziger:tn}
D.~Zwanziger,
Nucl.\ Phys.\ B {\bf 345}, 461 (1990).

\bibitem{Parrinello:1990pm}
C.~Parrinello and G.~Jona-Lasinio,
Phys.\ Lett.\ B {\bf 251}, 175 (1990).

\bibitem{Fachin:1991pu}
S.~Fachin and C.~Parrinello,
Phys.\ Rev.\ D {\bf 44}, 2558 (1991).

\bibitem{Giusti:2001xf}
L.~Giusti, M.~L.~Paciello, C.~Parrinello, S.~Petrarca and B.~Taglienti,
Int.\ J.\ Mod.\ Phys.\ A {\bf 16}, 3487 (2001)
[hep-lat/0104012].

\bibitem{Bock:2000cd}
W.~Bock, M.~Golterman, M.~Ogilvie and Y.~Shamir,
Phys.\ Rev.\ D {\bf 63}, 034504 (2001)
[hep-lat/0004017].

\bibitem{Henty:1996kv}
D.~S.~Henty,O.~Oliveira, C.~Parrinello and S.~Ryan [UKQCD Collaboration],
Phys.\ Rev.\ D {\bf 54}, 6923 (1996)
[hep-lat/9607014].

\bibitem{Fradkin:1978dv}
E.~H.~Fradkin and S.~H.~Shenker,
Phys.\ Rev.\ D {\bf 19}, 3682 (1979).

\bibitem{us_longer_ms}
C.~A.~Aubin and M.~C.~Ogilvie, in preparation.

\bibitem{Leinweber:1998uu}
D.~B.~Leinweber, J.~I.~Skullerud, A.~G.~Williams and C.~Parrinello [UKQCD
                  Collaboration]
Phys.\ Rev.\ D {\bf 60}, 094507 (1999)
[Erratum-ibid.\ D {\bf 61}, 079901 (2000)]
[hep-lat/9811027].

\bibitem{Teper:1998kw}
M.~J.~Teper,
[hep-th/9812187.]

\bibitem{Suzuki:1989gp}
T.~Suzuki and I.~Yotsuyanagi,
Phys.\ Rev.\ D {\bf 42}, 4257 (1990).

\bibitem{Hioki:1991ai}
S.~Hioki, S.~Kitahara, S.~Kiura, Y.~Matsubara, O.~Miyamura, S.~Ohno and T.~Suzuki,
Phys.\ Lett.\ B {\bf 272}, 326 (1991)
[Erratum-ibid.\ B {\bf 281}, 416 (1992)].

\bibitem{Mitrjushkin:2001hr}
V.~K.~Mitrjushkin and A.~I.~Veselov,
JETP Lett.\  {\bf 74}, 532 (2001)
[Pisma Zh.\ Eksp.\ Teor.\ Fiz.\  {\bf 74}, 605 (2001)]
[arXiv:hep-lat/0110200].


\bibitem{Faber:1998en}
M.~Faber, J.~Greensite and S.~Olejnik,
JHEP {\bf 9901}, 008 (1999)
[hep-lat/9810008].

\bibitem{Ogilvie:1998wu}
M.~C.~Ogilvie,
Phys.\ Rev.\ D {\bf 59}, 074505 (1999)
[hep-lat/9806018].

\bibitem{Stack:2000kf}
J.~D.~Stack and W.~W.~Tucker,
Nucl.\ Phys.\ Proc.\ Suppl.\  {\bf 94}, 529 (2001)
[arXiv:hep-lat/0011034].

\bibitem{Bornyakov:2000cd}
V.~G.~Bornyakov, D.~A.~Komarov, M.~I.~Polikarpov and A.~I.~Veselov,
JETP Lett.\  {\bf 71}, 231 (2000)
[arXiv:hep-lat/0002017].

\end{thebibliography}
\end{document}